\documentclass{IEEEmce}

\usepackage[colorlinks,urlcolor=blue,linkcolor=blue,citecolor=blue]{hyperref}
\usepackage[utf8]{inputenc}
\usepackage{url}

\usepackage{amsmath,amsfonts,bm}

\usepackage{graphicx}
\usepackage{adjustbox}
\usepackage{subcaption}
\usepackage{booktabs}
\usepackage{multirow}
\usepackage{array}
\usepackage{tabularx}
\newcommand{\cmark}{\ding{51}}
\usepackage{pifont}
\usepackage{makecell}
\usepackage{siunitx}

\usepackage{xcolor}
\usepackage{pifont}
\let\labelindent\relax
\usepackage{enumitem}
\usepackage{microtype}
\usepackage[skins,breakable,most]{tcolorbox}

\usepackage{tikz}
\usetikzlibrary{
  backgrounds,
  positioning,
  fit,
  calc,
  arrows.meta
}
\usepackage{pgfplots}
\usepgfplotslibrary{groupplots}
\pgfplotsset{compat=1.18}

\pgfplotsset{
  colormap={metricmap}{
    rgb255=(68,1,84)
    rgb255=(59,82,139)
    rgb255=(33,145,140)
    rgb255=(253,231,37)
  }
}

\jvol{XX}
\jnum{XX}
\paper{8}
\jmonth{xxx/xxx}
\publisheddate{00 xxxx 0000}
\currentdate{00 xxxx 0000}
\jname{IEEE Design \& Test}
\pubyear{2026}
\doiinfo{DNT.2026.Doi Number}

\begin{document}
\editor{Editor: Name, xxxx@email}
\bstctlcite{IEEEexample:BSTcontrol}
\title{GRADE-RTL: Evaluating LLM-Generated RTL Beyond Compilation}

\author{Hepziba Susan}
\affil{School of Electronics Engineering, Vellore Institute of Technology, Vellore, Tamil Nadu, India\\
hepziba.susan2022@vitstudent.ac.in}

\author{Shivaranjani G.~R.}
\affil{School of Electronics Engineering, Vellore Institute of Technology, Chennai, Tamil Nadu, India\\
shivaranjani.gr2023@vitstudent.ac.in}

\author{Malik Imran}
\affil{School of Electronics, Electrical Engineering and Computer Science, Queen's University Belfast, Belfast, United Kingdom\\
m.imran@qub.ac.uk}

\author{Muhammad Rashid}
\affil{Department of Computer and Network Engineering, College of Computing, Umm Al-Qura University, Makkah, Saudi Arabia\\
Email: mfelahi@uqu.edu.sa}

\author{Sumathi Gokulanathan}
\affil{School of Electronics Engineering, Vellore Institute of Technology, Vellore, Tamil Nadu, India\\
sumathi.g@vit.ac.in}

\author{Zain Ul Abideen}
\affil{Department of Electrical and Computer Engineering, University of Idaho, Moscow, ID, USA\\
zabideen@uidaho.edu}
\maketitle

\chapterinitial{Edge systems} combine control, communication, security, memory, and embedded-compute intellectual property (IP) under tight area, power, and timing constraints. At the same time, the growing complexity of these systems has increased the demand for automated design methodologies. In this context, Large Language Models (LLMs) are emerging as powerful design-time assistants capable of translating natural-language specifications into register-transfer-level (RTL) code. This capability could shorten the development of edge-hardware blocks, but only if the generated RTL is structurally valid, functionally correct, and efficient after implementation~\cite{pan2025survey}.

\begin{figure}[h!] 
\centering 
\includegraphics[width=\columnwidth]{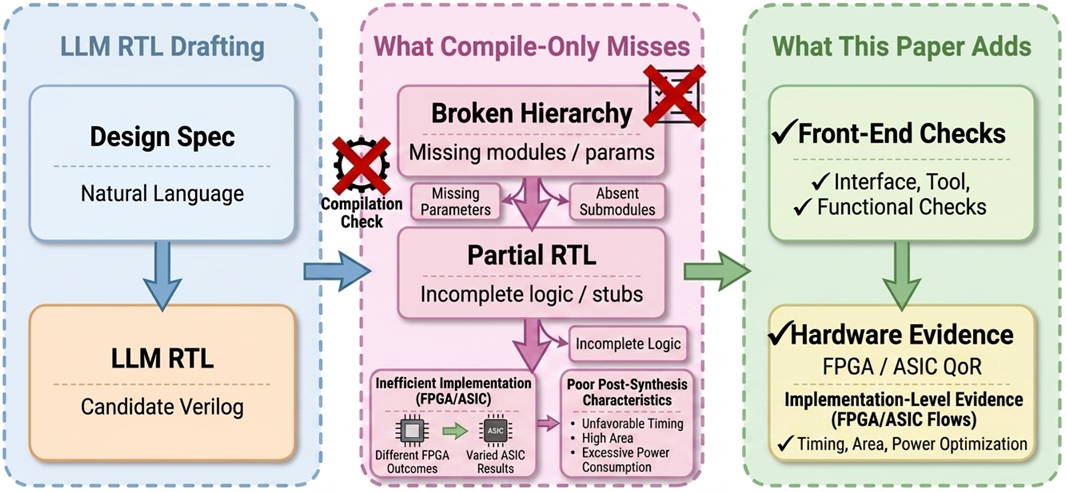} 
\caption{Motivation for implementation-aware RTL evaluation.}
\label{fig:gap} 
\end{figure}

To address these requirements, recent RTL-generation models and workflows have improved compilation success and functional correctness through domain-specific training and tool-driven feedback~\cite{thakur2023verigen,thakur2023autochip,liu2023chipnemo}. However, compilation success alone is insufficient to establish overall design quality. As illustrated in Fig.~\ref{fig:gap}, compilation-based evaluation provides an incomplete assessment of design quality. Critical issues, such as broken module hierarchies, missing parameters, incomplete logic, and absent submodules, may remain undetected despite successful compilation. 

In addition to these structural issues, compilation and functional correctness do not necessarily translate into efficient hardware implementations. Even functionally equivalent RTL descriptions can produce substantially different field-programmable gate array (FPGA) and application-specific integrated circuit (ASIC) implementations. Consequently, RTL that passes compilation and functional verification may still exhibit poor post-synthesis characteristics, such as unfavorable timing, area, or power consumption. Therefore, evaluation should extend beyond compilation metrics by incorporating front-end verification checks, such as interface and functional validation, together with implementation-level evidence from FPGA and ASIC design flows.

\begin{figure}[h!] 
\centering 
\includegraphics[width=0.92\columnwidth]{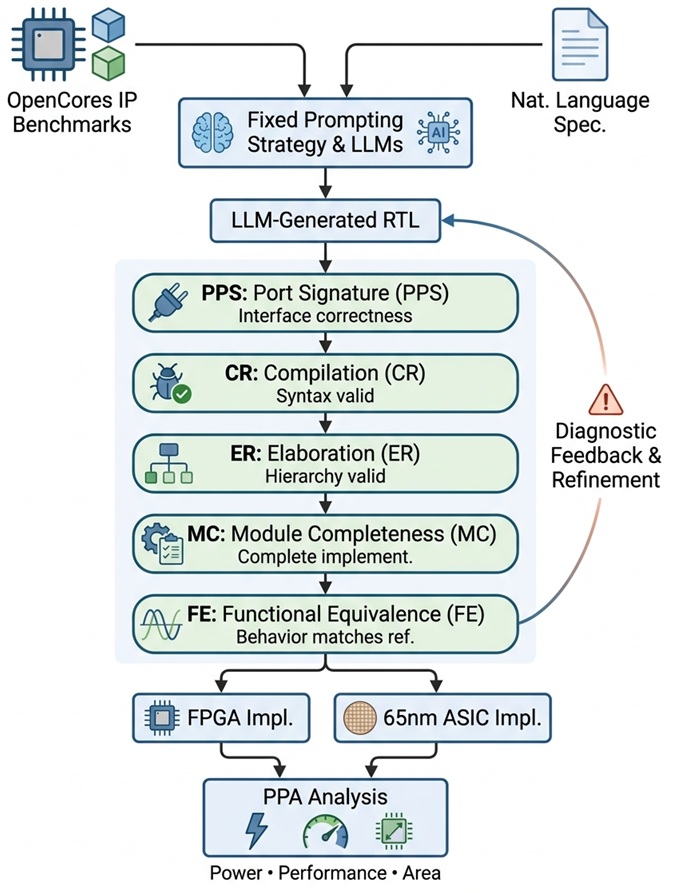} 
\caption{\textsc{GRADE-RTL}: Proposed five-stage evaluation of LLM-generated RTL.}
\label{fig:grade_rtl_framework} 
\end{figure}

Figure~\ref{fig:grade_rtl_framework} presents the proposed \textsc{GRADE-RTL} methodology for evaluating LLM-generated RTL. It employs edge-relevant OpenCores IP benchmarks and standardized natural-language specifications as inputs to multiple LLMs under a fixed prompting strategy. 
Subsequently, the generated RTL is assessed using a five-stage validation flow comprising Port Signature (PPS), Compilation (CR), Elaboration (ER), Module Completeness (MC), and Functional Equivalence (FE). Together, these stages verify interface correctness, syntactic validity, hierarchical consistency, implementation completeness, and behavioral equivalence with a trusted reference RTL design. 

To improve robustness, designs that fail validation may undergo bounded refinement using stage-specific diagnostic feedback generated from the earliest failed validation stage.  RTL candidates that satisfy the front-end validation stages are further evaluated using FPGA and 65\,nm ASIC implementation flows to quantify PPA characteristics. By linking front-end correctness with implementation-level QoR analysis, \textsc{GRADE-RTL} provides a comprehensive assessment of RTL generation quality beyond compilation-based metrics. To support reproducibility, all prompts, scripts, logs, model configurations, and evaluation artifacts are publicly released.\footnote{\label{fn:grade}\url{https://github.com/hsc-research/GRADE-RTL}}

\section{Related Work and Positioning}
\label{sec:background}

Domain-adapted generators, such as VeriGen~\cite{thakur2023verigen}, feedback-driven workflows such as AutoChip~\cite{thakur2023autochip}, and retrieval-augmented systems such as ChipNeMo~\cite{liu2023chipnemo} have improved RTL generation quality through specialized training and iterative feedback mechanisms. To assess these capabilities, benchmarks, such as VerilogEval~\cite{Mingjie2023} and RTLLM~\cite{lu2023rtllm} primarily evaluate compilation success and functional correctness. While these studies have advanced RTL generation research, they provide limited insight into design completeness, structural robustness, and the quality of generated RTL.

The closest study to our work is TuRTLe, which integrates multiple benchmarks and evaluates syntax, functional correctness, synthesizability, power, performance, area, and line-completion behavior across a diverse set of open-source models~\cite{garcia2025turtle}. We therefore position \textsc{GRADE-RTL} as a complementary framework rather than claiming the first unified RTL-evaluation methodology. Owing to differences in benchmark suites, model sets, protocols, and backend flows, direct numerical comparisons are not meaningful. Instead, the two studies provide complementary insights into RTL-generation quality and together offer a broader perspective on LLM capabilities for hardware design. Our paper differs from prior work in four respects: 

\begin{enumerate}[label=(\roman*),leftmargin=*] 

\item It evaluates realistic IP-level designs using trusted reference RTL obtained from OpenCores~\cite{anonymous_opencores_ip_benchmark}. 

\item It explicitly separates MC from FE, enabling a clear distinction between unfinished implementations and behavioral mismatches. 

\item It employs a fixed, failure-directed refinement budget for both open- and closed-source models. All models are evaluated under the same bounded interaction budget, ensuring that improvements reflect effective use of verification feedback rather than unrestricted prompt iteration, enabling fair and reproducible comparison.

\item It links front-end validation with FPGA and 65\,nm ASIC implementation evidence, including an Innovus place-and-route case study. This extends evaluation beyond RTL legality to assess implementation efficiency for edge-hardware IP.

\end{enumerate}

\section{Evaluation Framework} \label{sec:framework}
Fig.~\ref{fig:flow} illustrates the proposed evaluation framework. Stage 1 establishes a consistent evaluation environment across all model-design pairs. Each LLM is treated as a black-box RTL generator, while prompts, tool versions, validation scripts, and implementation constraints remain fixed. The accompanying artifact\footref{fn:grade} provides the complete framework, including generation scripts, benchmark prompts, refinement procedures, validation tools, and evaluation metrics. Both locally hosted and API-based LLMs are supported through configurable provider and model settings.

\tikzset{
  stage/.style={
    draw, dashed, rounded corners=10pt, very thick,
    inner xsep=10pt, inner ysep=12pt
  },
  sblock/.style={
    draw, rounded corners=6pt,
    minimum width=2.6cm, minimum height=1.6cm,
    align=center, font=\fontsize{9.5}{9.5}
  },
  sblue/.style={sblock, fill=blue!12},
  sorange/.style={sblock, fill=orange!15},
  sgreen/.style={sblock, fill=green!12},
  spink/.style={sblock, fill=magenta!12},
  syellow/.style={sblock, fill=yellow!25},
  stitle/.style={font=\bfseries\small, anchor=north}
}

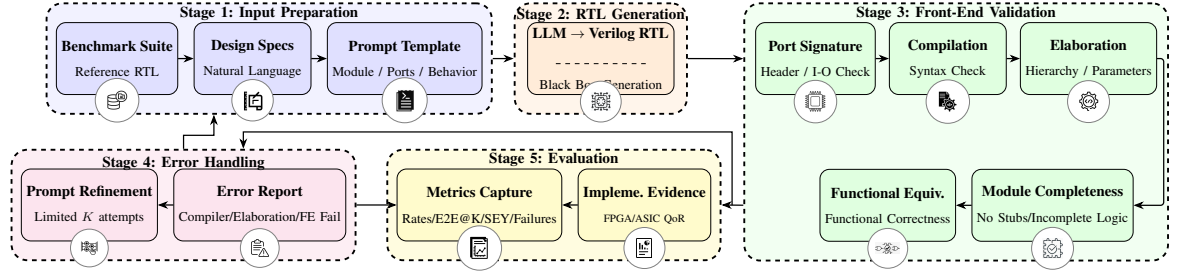
\begin{figure*}[h!]
\centering
\begin{adjustbox}{width=\textwidth}

\begin{tikzpicture}[>=Stealth, node distance=2mm and 2mm]


\node[sblue] (S1a) at (0,0)
{
\textbf{Benchmark Suite}\\[3pt]
{\small Reference RTL}
};

\node[sblue, right=3.5mm of S1a] (S1b)
{
\textbf{Design Specs}\\[3pt]
{\small Natural Language}
};

\node[sblue, right=3.5mm of S1b] (S1c)
{
\textbf{Prompt Template}\\[3pt]
{\small Module / Ports / Behavior}
};

\draw[->, thick] (S1a.east) -- (S1b.west);
\draw[->, thick] (S1b.east) -- (S1c.west);

\begin{scope}[on background layer]
\node[stage, inner xsep=6pt, inner ysep=12pt, fit=(S1a)(S1b)(S1c), fill=blue!5] (stage1) {};
\end{scope}

\node at ($(stage1.north)+(0,-2.5mm)$)
{\bfseries Stage 1: Input Preparation};

\node[circle, draw=gray, fill=white, minimum size=2em]
at ($(S1a.south)+(0,-1.2mm)$)
{\includegraphics[width=13pt]{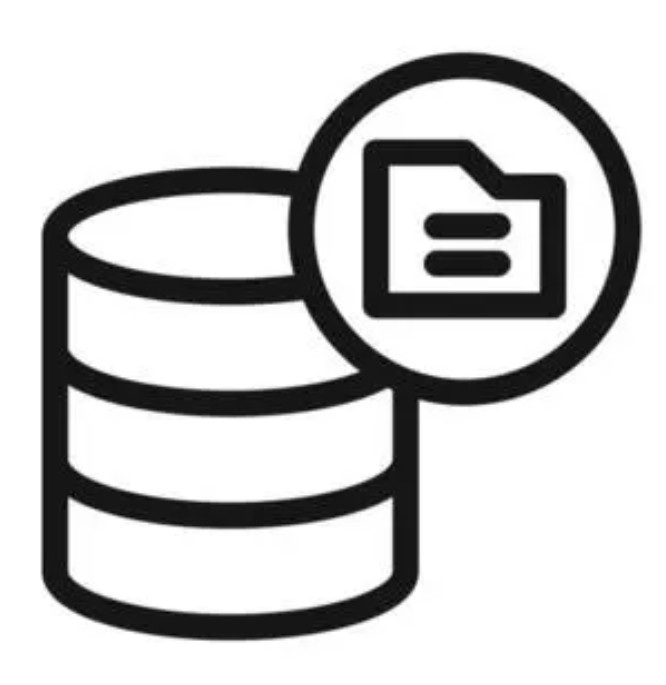}};

\node[circle, draw=gray, fill=white, minimum size=2em]
at ($(S1b.south)+(0,-1.2mm)$)
{\includegraphics[width=12pt]{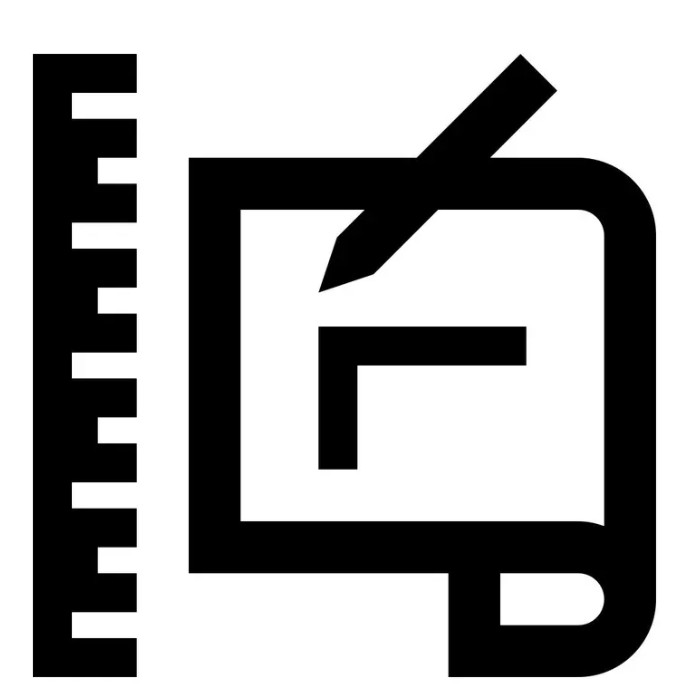}};

\node[circle, draw=gray, fill=white, minimum size=2em]
at ($(S1c.south)+(0,-1.2mm)$)
{\includegraphics[width=12pt]{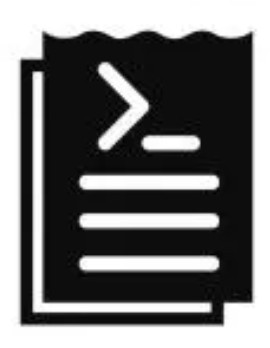}};


\node[sorange, right=10mm of S1c] (llm)
{
\textbf{LLM $\rightarrow$ Verilog RTL}\\[3pt]
\tikz[baseline]{\draw[dashed] (-1.0,0)--(1.0,0);}\\[3pt]
{\small Black Box Generation}
};

\begin{scope}[on background layer]
\node[stage, inner xsep=8pt, inner ysep=12pt,fit=(llm), fill=orange!8] (stage2) {};
\end{scope}

\node[stitle] at ($(stage2.north)+(0,-0.2mm)$)
{\normalsize Stage 2: RTL Generation};

\node[circle, draw=gray, fill=white, minimum size=2.2em]
at ($(llm.south)+(0,-1.5mm)$)
{\includegraphics[width=12pt]{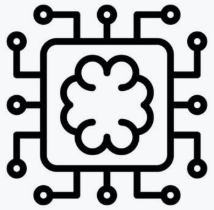}};


\node[sgreen, right=15mm of stage2] (pps)
{
\textbf{Port Signature}\\[3pt]
{\small Header / I-O Check}
};

\node[sgreen, right=3mm of pps] (comp)
{
\textbf{Compilation}\\[3pt]
{\small Syntax Check}
};

\node[sgreen, right=3mm of comp] (elab)
{
\textbf{Elaboration}\\[3pt]
{\small Hierarchy / Parameters}
};

\node[sgreen, xshift=-8mm, below=16.5mm of elab] (mc)
{
\textbf{Module Completeness}\\[3pt]
{\small No Stubs/Incomplete Logic}
};

\node[sgreen, left=3.5mm of mc] (fe)
{
\textbf{Functional Equiv.}\\[3pt]
{\small Functional Correctness}
};


\draw[->, thick] (pps.east) -- (comp.west);
\draw[->, thick] (comp.east) -- (elab.west);
\draw[->, thick, rounded corners=2pt] (elab.east) -- ++(1.5mm,0) |- (mc.east);
\draw[->, thick] (mc.west) -- (fe.east);

\begin{scope}[on background layer]
\node[stage, inner xsep=7pt, inner ysep=12pt, fit=(pps)(comp)(elab)(mc)(fe), fill=green!5] (stage3) {};
\end{scope}

\node[stitle] at ($(stage3.north)+(0,0.2mm)$)
{\normalsize Stage 3: Front-End Validation};

\node[circle, draw=gray, fill=white, minimum size=2.2em]
at ($(pps.south)+(0,-1.5mm)$)
{\includegraphics[width=12.5pt]{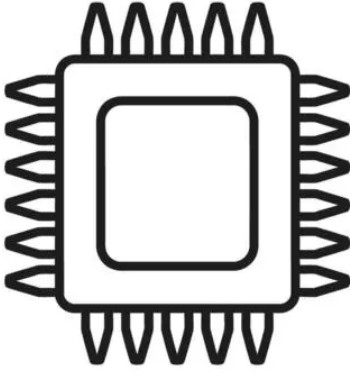}};

\node[circle, draw=gray, fill=white, minimum size=2.2em]
at ($(comp.south)+(0,-1.5mm)$)
{\includegraphics[width=12.5pt]{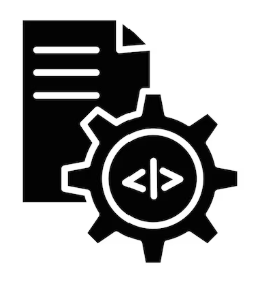}};

\node[circle, draw=gray, fill=white, minimum size=2.2em]
at ($(elab.south)+(0,-1.5mm)$)
{\includegraphics[width=12.5pt]{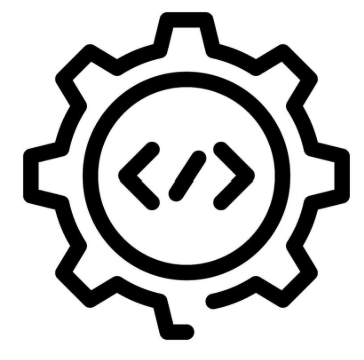}};

\node[circle, draw=gray, fill=white, minimum size=2.2em]
at ($(mc.south)+(0,-1.5mm)$)
{\includegraphics[width=12.5pt]{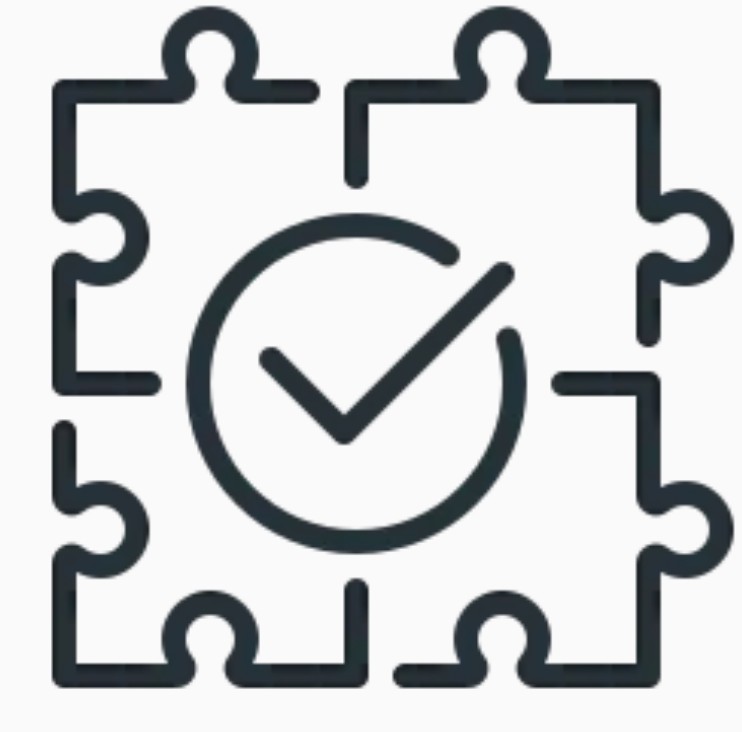}};

\node[circle, draw=gray, fill=white, minimum size=2.2em]
at ($(fe.south)+(0,-1.5mm)$)
{\includegraphics[width=15pt]{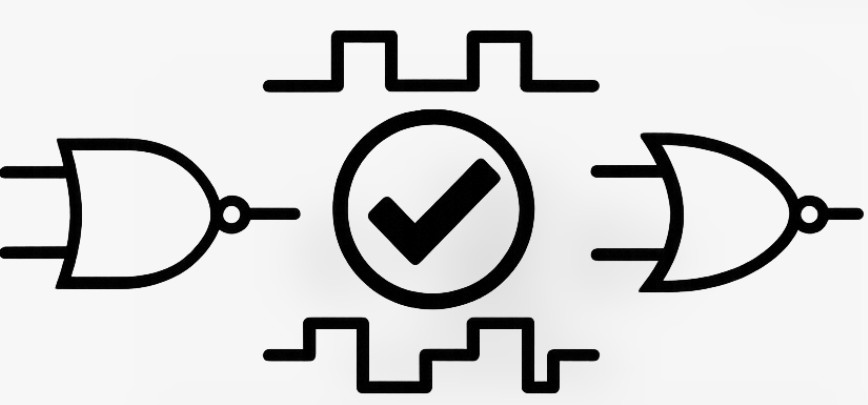}};



\node[spink, below=12mm of stage1, xshift=-40mm] (ref)
{
\textbf{Prompt Refinement}\\[3pt]
{\small Limited $K$ attempts}
};

\node[spink, right=3.5mm of ref] (err)
{
\textbf{Error Report}\\[3pt]
{\small Compiler/Elaboration/FE Fail}
};

\draw[->, thick] (err.west) -- (ref.east);

\begin{scope}[on background layer]
\node[stage, inner xsep=6pt, inner ysep=12pt, fit=(ref)(err), fill=magenta!8] (stage4) {};
\end{scope}

\node[stitle] at ($(stage4.north)+(0,-0.5mm)$)
{\normalsize Stage 4: Error Handling};

\node[circle, draw=gray, fill=white, minimum size=2.2em]
at ($(ref.south)+(0,-1.5mm)$)
{\includegraphics[width=11pt]{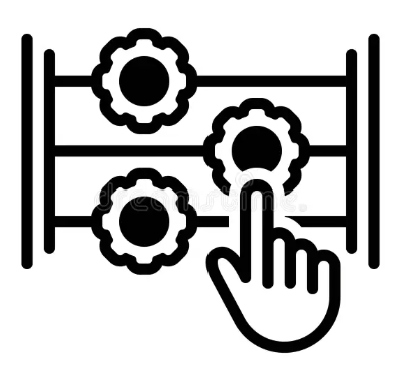}};

\node[circle, draw=gray, fill=white, minimum size=2.2em]
at ($(err.south)+(0,-1.5mm)$)
{\includegraphics[width=11pt]{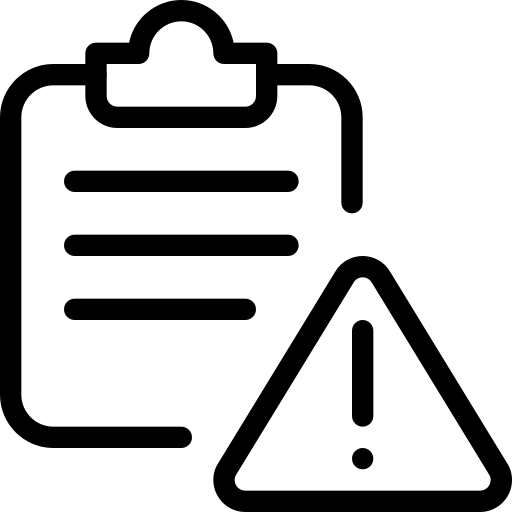}};


\node[syellow, below=12mm of stage2, xshift=-27mm] (metric)
{
\textbf{Metrics Capture}\\[3pt]
{\small Rates/E2E@K/SEY/Failures}
};

\node[syellow, right=3.5mm of metric] (score)
{
\textbf{Impleme. Evidence}\\[3pt]
{\scriptsize FPGA/ASIC QoR}
};

\draw[->, thick] (score.west) -- (metric.east);

\begin{scope}[on background layer]
\node[stage, inner xsep=6pt, inner ysep=12pt, fit=(metric)(score), fill=yellow!15] (stage5) {};
\end{scope}

\node at ($(stage5.north)+(0,-2.4mm)$)
{\bfseries Stage 5: Evaluation};

\node[circle, draw=gray, fill=white, minimum size=2.2em]
at ($(metric.south)+(0,-1.5mm)$)
{\includegraphics[width=12pt]{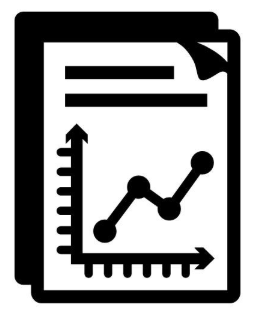}};

\node[circle, draw=gray, fill=white, minimum size=2.2em]
at ($(score.south)+(0,-1.5mm)$)
{\includegraphics[width=12pt]{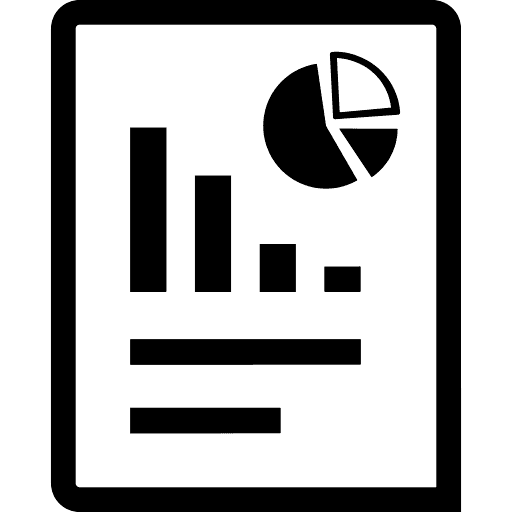}};


\draw[->, thick] (stage1.east) -- (stage2.west);
\draw[->, thick] (stage2.east) -- ([yshift=16.3mm]stage3.west);
\draw[->, thick] ([yshift=-16.3mm]stage3.west) -- (stage5.east);
\draw[->, thick] (stage4.east) -- (metric.west);

\draw[->, thick] ([yshift=-16.3mm, xshift=-2.5mm]stage3.west) -- ++(0,16.3mm) -| ([xshift=13.3mm]stage4.north);


\draw[->, thick] (stage4.north) -- ++(0,3mm) -| ([xshift=-12.3mm]stage1.south);

\end{tikzpicture}

\end{adjustbox}
\vspace{-3mm}
\caption{The \textsc{GRADE-RTL: } End-to-end evaluation framework.}
\label{fig:flow}
\end{figure*}

In Stage 2 of Fig.~\ref{fig:flow}, the benchmark suite provides ten OpenCores-based designs spanning control, communication, security, error correction, arithmetic, processing, and memory functions~\cite{anonymous_opencores_ip_benchmark}. Each design includes a trusted reference RTL and a natural-language specification. Unlike existing benchmarks that mainly focus on small circuits and educational designs~\cite{Mingjie2023,lu2023rtllm}, our suite emphasizes realistic multi-module IP blocks with diverse interfaces, control logic, datapaths, and hierarchies. The benchmark prioritizes architectural diversity while maintaining a reproducible evaluation flow. For each design, a standardized prompt is constructed from the interface, reset convention, behavioral requirements, and synthesizability constraints before being provided to the LLM.

In Stage 3 of Fig.~\ref{fig:flow}, generated RTL undergoes a five-stage front-end validation flow. Port Signature (PPS) verifies the module interface, while Compilation (CR) checks Verilog-2001 syntax and legality. Elaboration (ER) resolves hierarchy, parameters, and bindings. Module Completeness (MC) identifies placeholders, undriven outputs, and incomplete logic. Finally, Functional Equivalence (FE) compares the generated RTL against a trusted reference design to verify behavioral correctness

To improve robustness, the framework supports bounded refinement in stage 4 (error handling). When a design fails validation, feedback is generated from the earliest failing stage. The model receives a concise diagnostic message describing the error, such as syntax issues, parameter mismatches, reset errors, or missing modules. It does not have access to the validation tools, logs, or implementation reports. A fixed refinement budget of $K=3$ is applied to all models and benchmarks. This controlled feedback process enables fair and reproducible evaluation of a model's ability to utilize verification feedback.

In Stage 5, FE is performed using a Yosys-based equivalence-checking flow that compares the generated RTL against a trusted reference design. Designs are classified as FE failures if equivalence cannot be established or if verification ends in an error or timeout. The framework reports stage-wise success rates, conditional yields, E2E@1, E2E@K, SEY@K, and first-failure breakdowns. Candidates that pass front-end validation are further evaluated using FPGA and 65\,nm ASIC implementation flows to quantify power, performance, and area (PPA) characteristics.

\section{Evaluation Metrics}
\label{sec:metrics}

While the five-stage validation flow in previous section (Fig.~\ref{fig:flow}) identifies where RTL generation succeeds or fails, quantitative metrics are needed for consistent model comparison and refinement analysis. The proposed scorecard measures progression through the validation pipeline, one-shot success, refinement recovery, and synthesis survivability. Stage-wise rates and conditional yields further localize failures across PPS, CR, ER, MC, and FE.

Let \(N\) denote the number of benchmark designs, and let \(y_i^{(a)}=1\) if design \(i\) pass all five validation stages on attempt \(a\); otherwise, \(y_i^{(a)}=0\). The end-to-end success rate on the first attempt is defined as: 

\[ \mathrm{E2E@1} = \frac{1}{N} \sum_{i=1}^{N} y_i^{(1)}. \] 

The end-to-end success rate within a refinement budget of \(K\) attempts is: 
\[ \mathrm{E2E@}K = \frac{1}{N} \sum_{i=1}^{N} \mathbb{1} \left\{ \max_{1 \le a \le K} y_i^{(a)} =1 \right\}, \] where \(\mathbb{1}\{\cdot\}\) is the indicator function.
The difference between the two metrics quantifies the effectiveness of verification-guided refinement. 

In addition to correctness, the framework measures implementation survivability through the synthesis-eligible yield, SEY@\(K\), defined as the fraction of designs that pass CR and ER within the refinement budget. Unlike E2E@\(K\), SEY@\(K\) reflects survivability rather than correctness. Thus, a gap between the two metrics identifies designs that can enter synthesis but remain incomplete or functionally incorrect. We also report first-failure breakdowns, categorizing failures as interface/compiler, non-elaboration, partial-module, or functional-mismatch failures.

Consistent with Stage~5 of Fig.~\ref{fig:flow}, implementation-level QoR analysis is performed only for candidates that pass both MC and FE. These candidates proceed to FPGA and 65\,nm ASIC implementation flows for PPA evaluation. Designs that satisfy synthesis eligibility alone are retained in survivability statistics but are not treated as correct implementations.

\section{Results}
\label{sec:results}

This section evaluates \textsc{GRADE-RTL} on nine LLMs and ten edge-relevant RTL designs. All models use the same prompt structure, reasonable bounded refinement budget (\(K=3\)), validation scripts, and FPGA/ASIC implementation settings. The evaluated set includes RTL-specialized models (OriGen, RTLCoder, VeriGen, VeriSeek, and CodeV~\cite{Fan2025,Shang2025,thakur2023verigen,VeriSeek2025,CodeV2024}) as well as general-purpose LLMs.

The results are presented from four complementary perspectives. We first analyze front-end robustness across the validation pipeline, followed by FPGA and ASIC implementation results for functionally correct designs. A PID-controller case study then highlights implementation-level differences, and the section concludes with a summary of overall model behavior using E2E@1, E2E@\(K\), SEY@\(K\), and first-failure metrics.



\subsection{Front-End Robustness}

Table~\ref{tab:coverage} records the within-budget outcome for every
model-design pair. A checkmark denotes end-to-end success within
\(K=3\); otherwise, the entry gives the first failing stage of the
selected deepest-progressing attempt: Port Signature (P), Compilation
(C), Elaboration (E), Module Completeness (M), or Functional
Equivalence (F).

Early-stage legality does not guarantee complete RTL. Several models
pass PPS and Compilation but lose candidates at Elaboration or Module
Completeness. Difficulty is also design-specific rather than determined solely by category: AES-128, SHA-256, and Toom--Cook are among the least frequently solved designs, whereas Present Cipher and Schoolbook are solved by several models.

All candidates that pass Compilation and Elaboration contribute to
SEY@\(\!K\). To avoid conflating synthesizability with correctness, the comparative FPGA and ASIC QoR results below include only candidates that also pass MC and FE. Synthesis-eligible candidates that fail a later stage remain in the survivability statistics but are not treated as correct implementations.

\begingroup
\setlength{\tabcolsep}{3pt}
\begin{table}[htbp]
\footnotesize\centering
\caption{Evaluation summary of the designs within $K=3$}
\label{tab:coverage}
\begin{tabular}{p{2.0cm}p{0.4cm}p{0.4cm}p{0.4cm}p{0.4cm}p{0.4cm}p{0.4cm}p{0.4cm}p{0.4cm}p{0.4cm}}
\toprule
\textbf{Design} &
\rotatebox[origin=c]{90}{\textbf{OriGen~\cite{Fan2025}}} &
\rotatebox[origin=c]{90}{\textbf{RTLCoder~\cite{Shang2025}}} &
\rotatebox[origin=c]{90}{\textbf{VeriGen~\cite{thakur2023verigen}}} &
\rotatebox[origin=c]{90}{\textbf{VeriSeek~\cite{VeriSeek2025}}} &
\rotatebox[origin=c]{90}{\textbf{CodeV~\cite{CodeV2024}}} &
\rotatebox[origin=c]{90}{\textbf{Gemini}} &
\rotatebox[origin=c]{90}{\textbf{DeepSeek}} &
\rotatebox[origin=c]{90}{\textbf{GPT-5}} &
\rotatebox[origin=c]{90}{\textbf{Claude}} \\

\midrule
Reed Solomon      & P & C & P & P & F & M & E & E & \cmark \\
PID Controller    & M & \cmark & C & C & C & M & E & M & \cmark \\
AES-128           & P & P & P & P & P & M & E & E & \cmark \\
Present Cipher    & P & P & P & P & F & \cmark & \cmark & \cmark & \cmark \\
SHA-256           & P & P & P & P & P & E & E & M & M \\
Toom-Cook         & F & C & P & C & P & M & E & M & M \\
Schoolbook        & \cmark & P & C & E & \cmark & \cmark & E & M & \cmark \\
UART 16550        & E & C & P & C & P & M & E & \cmark & \cmark \\
uRISC-V           & F & E & E & P & M & M & \cmark & E & E \\
Wishbone RAM  & P & C & P & P & P & M & \cmark & E & \cmark \\
\bottomrule
\end{tabular}
\end{table}
\endgroup

\begin{figure*}[htb]
\centering
\begin{tikzpicture}
\begin{groupplot}[
    group style={group size=4 by 1, horizontal sep=0.85cm},
    width=4.5cm,
    height=3.30cm,
    xmin=0.5, xmax=6.5,
    xtick={1,2,3,4,5,6},
    xticklabels={PID Controller,Present Cipher,SHA-256,Schoolbook,UART 16550,Wishbone RAM},
    xticklabel style={rotate=40, anchor=east, font=\scriptsize},
    tick label style={font=\scriptsize},
    label style={font=\scriptsize},
    title style={font=\scriptsize, yshift=-1mm},
    grid=none
]

\nextgroupplot[
    title={LUT Overhead},
    ylabel={\(\times\) baseline},
    title style={yshift=-3pt}, 
    ymin=0, ymax=3.2,
    ytick={0,1,2,3},
    ylabel style={yshift=-3pt},
  legend style={font=\scriptsize, at={(2.35,1.47)}, anchor=north,
  legend columns=4, inner sep=1pt, row sep=2pt, column sep=3pt,
  nodes={scale=1.0, transform shape},
  legend image post style={scale=0.7}}
]
\addplot[dashed, black, forget plot] coordinates {(0.5,1) (6.5,1)};

\addplot[only marks, mark=*, mark size=2.6pt, blue] coordinates {
(1,0.673) (2,1.028) (5,0.358) (6,0.863)
};
\addlegendentry{Claude}

\addplot[only marks, mark=square*, mark size=2.6pt, orange!80!black] coordinates {
(1,0.406) (2,1.145) (4,0.525)
};
\addlegendentry{Gemini}

\addplot[only marks, mark=triangle*, mark size=2.9pt, red] coordinates {
(3,2.915)
};
\addlegendentry{GPT-5}

\addplot[only marks, mark=diamond*, mark size=2.9pt, purple] coordinates {
(4,1.361)
};
\addlegendentry{OriGen}

\nextgroupplot[
    title={FF Overhead},
    ylabel={\(\times\) baseline},
    title style={yshift=-3pt}, 
    ymin=0, ymax=3.2,
    ytick={0,1,2,3},
    ylabel style={yshift=-3pt}
]
\addplot[dashed, black, forget plot] coordinates {(0.5,1) (6.5,1)};

\addplot[only marks, mark=*, mark size=2.6pt, blue] coordinates {
(1,1.652) (2,1.007) (5,0.514) (6,1.249)
};

\addplot[only marks, mark=square*, mark size=2.6pt, orange!80!black] coordinates {
(1,0.836) (2,1.416) (4,2.933)
};

\addplot[only marks, mark=triangle*, mark size=2.9pt, red] coordinates {
(3,2.479)
};

\addplot[only marks, mark=diamond*, mark size=2.9pt, purple] coordinates {
(4,2.904)
};

\nextgroupplot[
    title={Power Overhead},
    ylabel={\(\times\) baseline},
    title style={yshift=-3pt}, 
    ymin=0.6, ymax=1.2,
    ytick={0.6,0.8,1.0,1.2},
    ylabel style={yshift=-3.5pt}
]
\addplot[dashed, black, forget plot] coordinates {(0.5,1) (6.5,1)};

\addplot[only marks, mark=*, mark size=2.6pt, blue] coordinates {
(1,1.064) (2,1.053) (5,0.993) (6,1.019)
};

\addplot[only marks, mark=square*, mark size=2.6pt, orange!80!black] coordinates {
(1,1.028) (2,0.820) (4,0.724)
};

\addplot[only marks, mark=triangle*, mark size=2.9pt, red] coordinates {
(3,0.841)
};

\addplot[only marks, mark=diamond*, mark size=2.9pt, purple] coordinates {
(4,0.724)
};

\nextgroupplot[
    title={Slack Difference},
    title style={yshift=-3pt}, 
    ylabel={ns},
    ymin=-4.5, ymax=4.5,
    ytick={-4,-2,0,2,4},
    ylabel style={yshift=-6pt}
]
\addplot[dashed, black, forget plot] coordinates {(0.5,0) (6.5,0)};

\addplot[only marks, mark=*, mark size=2.6pt, blue] coordinates {
(1,-1.305) (2,-0.112) (5,1.280) (6,2.381)
};

\addplot[only marks, mark=square*, mark size=2.6pt, orange!80!black] coordinates {
(2,-1.688) (4,1.640)
};

\addplot[only marks, mark=triangle*, mark size=2.9pt, red] coordinates {
(3,-0.485)
};

\addplot[only marks, mark=diamond*, mark size=2.9pt, purple] coordinates {
(4,-0.228)
};

\end{groupplot}

\end{tikzpicture}
\caption{Normalized FPGA implementation overheads relative to the baseline RTL for representative synthesis-eligible designs.}
\label{fig:fpga_norm_qor}
\vspace{-4pt}
\end{figure*}
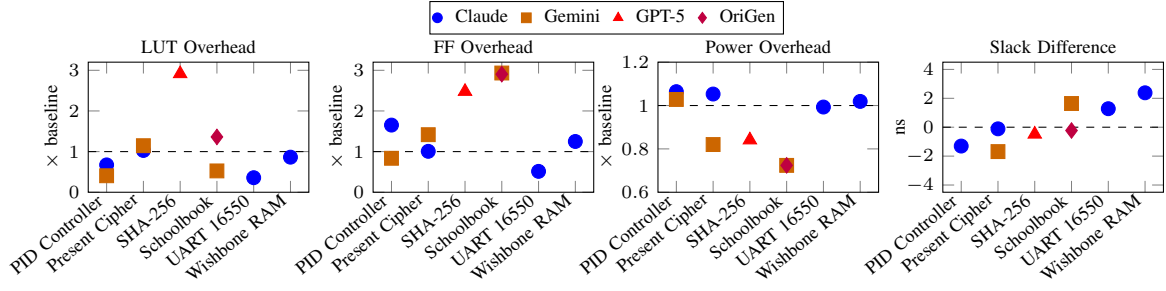

\subsection{FPGA Implementation}
\label{fpga}

Fig.~\ref{fig:fpga_norm_qor} compares representative FE-passing RTL
against the corresponding reference implementations. FPGA implementations use AMD Vivado targeting Artix-7 with a 100-MHz clock and identical synthesis, placement, routing, and I/O constraints. A value of one indicates baseline resource or power use, while the slack panel reports the absolute difference from the baseline at 100\,MHz.

Implementation overhead is strongly design-dependent. Control and
communication blocks, such as UART and Wishbone RAM, remain
comparatively close to their references. The Schoolbook multiplier
shows a wider spread: Gemini and OriGen use substantially more
flip-flops than the baseline, whereas Claude remains closer in resource
use but differs in timing. Present Cipher also exhibits model-dependent
resource and power trade-offs.

The result is not that generated RTL is uniformly worse. Instead,
functionally correct descriptions can expose different resource-sharing,
pipelining, and combinational-logic choices. FPGA QoR therefore provides
an implementation lens that functional equivalence alone cannot supply.

\subsection{ASIC Synthesis}
\label{asic}

\begin{table}[!ht]
\centering
\footnotesize
\setlength{\tabcolsep}{3pt}
\renewcommand{\arraystretch}{1.12}
\caption{Baseline and representative LLM ASIC implementations using
Cadence Genus at TT, 1.0\,V, and 125\,\(^\circ\)C.}
\label{tab:asic_results_merged}

\begin{tabular}{
p{1.25cm}
p{1.00cm}
>{\centering\arraybackslash}p{0.62cm}
>{\centering\arraybackslash}p{0.82cm}
>{\centering\arraybackslash}p{1.05cm}
>{\centering\arraybackslash}p{0.78cm}
}
\toprule
\textbf{Design} &
\textbf{Model} &
\multicolumn{1}{c}{\textbf{Timing}} &
\multicolumn{2}{c}{\textbf{Area}} &
\multicolumn{1}{c}{\textbf{Power}} \\
\cmidrule(lr){3-3}
\cmidrule(lr){4-5}
\cmidrule(lr){6-6}
& &
\makecell{\textbf{Slack}\\\textbf{(ns)}} &
\textbf{Cells} &
\makecell{\textbf{Area}\\\textbf{($\mu$m$^2$)}} &
\makecell{\textbf{Total}\\\textbf{(mW)}} \\
\midrule


\multirow{3}{*}{\shortstack[l]{PID\\Controller}}
& Baseline & 6.70 & 1,176 & 4,828.0 & 0.424 \\
& RTLCoder & 6.76 & 899 & 4,329.0 & 0.138 \\
& Claude & 6.76 & 6,794 & 27,382.3 & 1.284 \\
\midrule


\multirow{4}{*}{\shortstack[l]{Present\\Cipher}}
& Baseline & 9.04 & 1,041 & 2,963.5 & 3.456 \\
& Gemini & 8.76 & 807 & 2,591.0 & 0.206 \\
& Claude & 8.57 & 929 & 2,980.0 & 0.271 \\
& GPT-5 & 9.05 & 817 & 2,712.6 & 0.265 \\
\midrule



\multirow{4}{*}{Schoolbook}
& Baseline & 2.35 & 3,025 & 7,384.7 & 2.619 \\
& OriGen & 8.67 & 2,842 & 11,127.6 & 4.991 \\
& Gemini & 7.90 & 878 & 5,322.2 & 0.575 \\
& Claude & 0.41 & 1,969 & 10,183.6 & 0.931 \\
\midrule

\multirow{3}{*}{\shortstack[l]{UART\\16550}}
& Baseline & 7.54 & 1,422 & 5,734.1 & 0.543 \\
& Claude & 8.31 & 1,346 & 5,508.7 & 0.385 \\
& GPT-5 & 8.71 & 902 & 3,034.6 & 0.301 \\
\midrule


\multirow{2}{*}{\shortstack[l]{Wishbone\\RAM}}
& Baseline & 6.34 & 3,663 & 14,330.9 & 1.315 \\
& Claude & 8.33 & 3,058 & 17,190.0 & 1.006 \\
\bottomrule
\end{tabular}

\vspace{2pt}
\parbox{\columnwidth}{\footnotesize\textit{Note: Total power is the
sum of the reported static and dynamic power components. Slack is
measured at a 100\,MHz target frequency.}}
\end{table}

Table~\ref{tab:asic_results_merged} reports representative FE-passing
candidates synthesized with the same 65\,nm library, PVT corner, and
constraints. The fixed environment isolates differences arising from
RTL organization rather than technology or script variation.

Control-oriented designs are relatively stable. The UART candidates use
fewer cells than the baseline while maintaining positive slack.
Arithmetic blocks exhibit wider variation. For Schoolbook, Gemini is comparatively compact, OriGen increases area and total power, and Claude remains functionally correct but operates
much closer to the timing boundary. The PID candidates span both a
compact RTLCoder implementation and a substantially expanded Claude
implementation despite similar positive slack.

These results are logic synthesis-level QoR evidence, not physical sign-off.
Their role is to show that equivalent RTL descriptions can expose very
different optimization opportunities before place-and-route.

\begin{figure*}[t]
\centering
\begin{adjustbox}{max width=0.98\textwidth,center}
\begin{tikzpicture}

\begin{groupplot}[
    group style={
      group size=4 by 1,
      horizontal sep=1.30cm
    },
    height=3.20cm,
    scale only axis=true,
    tick label style={font=\small},
    label style={font=\small},
    title style={font=\small, yshift=1mm},
    xticklabel style={
      rotate=45,
      anchor=east,
      font=\small
    },
    grid=none
]

\nextgroupplot[
    width=3.95cm,
    title={Stage-wise Rates},
    xlabel={Model},
    ylabel={Stage},
    xmin=-0.5, xmax=8.5,
    ymin=-0.5, ymax=4.5,
    enlargelimits=false,
    xtick={0,1,2,3,4,5,6,7,8},
    xticklabels={
      OriGen,
      RTLCoder,
      VeriGen,
      VeriSeek,
      CodeV,
      Gemini,
      DeepSeek,
      GPT-5,
      Claude
    },
    ytick={0,1,2,3,4},
    yticklabels={PPS,CR,ER,MC,FE},
    y dir=reverse,
    tick style={draw=none},
    point meta min=0,
    point meta max=1,
    colormap name=metricmap
]

\addplot[
    matrix plot*,
    mesh/cols=9,
    point meta=explicit,
    draw=white,
    line width=0.25pt
]
table[meta=val] {
x y val
0 0 0.50
1 0 0.60
2 0 0.30
3 0 0.40
4 0 0.50
5 0 1.00
6 0 1.00
7 0 1.00
8 0 1.00
0 1 0.50
1 1 0.20
2 1 0.10
3 1 0.10
4 1 0.40
5 1 1.00
6 1 1.00
7 1 1.00
8 1 1.00
0 2 0.40
1 2 0.10
2 2 0.00
3 2 0.00
4 2 0.40
5 2 0.90
6 2 0.30
7 2 0.60
8 2 0.90
0 3 0.30
1 3 0.10
2 3 0.00
3 3 0.00
4 3 0.30
5 3 0.20
6 3 0.30
7 3 0.20
8 3 0.70
0 4 0.10
1 4 0.10
2 4 0.00
3 4 0.00
4 4 0.10
5 4 0.20
6 4 0.30
7 4 0.20
8 4 0.70
};

\nextgroupplot[
    width=3.95cm,
    title={Conditional Yields},
    xlabel={Model},
    ylabel={Yield},
    xmin=-0.5, xmax=8.5,
    ymin=-0.5, ymax=3.5,
    enlargelimits=false,
    xtick={0,1,2,3,4,5,6,7,8},
    xticklabels={
      OriGen,
      RTLCoder,
      VeriGen,
      VeriSeek,
      CodeV,
      Gemini,
      DeepSeek,
      GPT-5,
      Claude
    },
    ytick={0,1,2,3},
    yticklabels={C$|$P,E$|$C,M$|$E,F$|$M},
    y dir=reverse,
    tick style={draw=none},
    point meta min=0,
    point meta max=1,
    colormap name=metricmap,
    colorbar,
    colorbar style={
      width=1.4mm,
      height=2.15cm,
      ytick={0,0.5,1},
      yticklabel style={font=\small},
      ytick style={draw=none},
      at={(1.03,0.5)},
      anchor=west
    }
]

\addplot[
    matrix plot*,
    mesh/cols=9,
    point meta=explicit,
    draw=white,
    line width=0.25pt
]
table[meta=val] {
x y val
0 0 1.00
1 0 0.33
2 0 0.33
3 0 0.25
4 0 0.80
5 0 1.00
6 0 1.00
7 0 1.00
8 0 1.00
0 1 0.80
1 1 0.50
2 1 0.00
3 1 0.00
4 1 1.00
5 1 0.90
6 1 0.30
7 1 0.60
8 1 0.90
0 2 0.75
1 2 1.00
2 2 0.00
3 2 0.00
4 2 0.75
5 2 0.22
6 2 1.00
7 2 0.33
8 2 0.78
0 3 0.33
1 3 1.00
2 3 0.00
3 3 0.00
4 3 0.33
5 3 1.00
6 3 1.00
7 3 1.00
8 3 1.00
};

\nextgroupplot[
    width=3.95cm,
    xshift=5mm,
    title={Success and Eligibility},
    xlabel={Model},
    ylabel={Rate},
    symbolic x coords={
      OriGen,
      RTLCoder,
      VeriGen,
      VeriSeek,
      CodeV,
      Gemini,
      DeepSeek,
      GPT-5,
      Claude
    },
    xtick=data,
    ymin=0,
    ymax=1.05,
    ytick={0,0.25,0.5,0.75,1},
    enlarge x limits=0.12,
    legend style={
      font=\small,
      at={(0.5,1.24)},
      anchor=south,
      legend columns=3,
      inner sep=0.2pt,
      row sep=2pt,
      column sep=1.5pt,
      nodes={scale=1.0,transform shape},
      legend image code/.code={
        \draw[##1]
          (0cm,-0.06cm)
          rectangle
          (0.18cm,0.06cm);
      }
    }
]

\addplot+[
    ybar,
    bar width=2.7pt,
    bar shift=-3pt,
    fill=purple!35,
    draw=black,
    mark=none
]
coordinates {
(OriGen,0.10)
(RTLCoder,0.00)
(VeriGen,0.00)
(VeriSeek,0.00)
(CodeV,0.10)
(Gemini,0.10)
(DeepSeek,0.20)
(GPT-5,0.10)
(Claude,0.60)
};

\addplot+[
    ybar,
    bar width=2.7pt,
    bar shift=0pt,
    fill=blue!55,
    draw=black,
    mark=none
]
coordinates {
(OriGen,0.10)
(RTLCoder,0.10)
(VeriGen,0.00)
(VeriSeek,0.00)
(CodeV,0.10)
(Gemini,0.20)
(DeepSeek,0.30)
(GPT-5,0.20)
(Claude,0.70)
};

\addplot+[
    ybar,
    bar width=2.7pt,
    bar shift=3pt,
    fill=teal!65!black,
    draw=black,
    mark=none
]
coordinates {
(OriGen,0.40)
(RTLCoder,0.10)
(VeriGen,0.00)
(VeriSeek,0.00)
(CodeV,0.40)
(Gemini,0.90)
(DeepSeek,0.30)
(GPT-5,0.60)
(Claude,0.90)
};

\legend{E2E@1,E2E@$K$,SEY@$K$}

\nextgroupplot[
    width=3.95cm,
    xshift=5mm,
    title={First-Failure Breakdown},
    xlabel={Model},
    ylabel={Fraction},
    symbolic x coords={
      OriGen,
      RTLCoder,
      VeriGen,
      VeriSeek,
      CodeV,
      Gemini,
      DeepSeek,
      GPT-5,
      Claude
    },
    xtick=data,
    ymin=0,
    ymax=1.05,
    ytick={0,0.25,0.5,0.75,1},
    ybar stacked,
    bar width=4pt,
    enlarge x limits=0.10,
    legend style={
      font=\small,
      at={(0.5,1.24)},
      anchor=south,
      legend columns=4,
      inner sep=0.2pt,
      row sep=2pt,
      column sep=1.5pt,
      nodes={scale=1.0,transform shape},
      legend image code/.code={
        \draw[##1]
          (0cm,-0.06cm)
          rectangle
          (0.18cm,0.06cm);
      }
    }
]

\addplot+[
    fill=purple!55,
    draw=black
]
coordinates {
(OriGen,0.556)
(RTLCoder,0.889)
(VeriGen,0.900)
(VeriSeek,0.900)
(CodeV,0.667)
(Gemini,0.000)
(DeepSeek,0.000)
(GPT-5,0.000)
(Claude,0.000)
};

\addplot+[
    fill=blue!55,
    draw=black
]
coordinates {
(OriGen,0.111)
(RTLCoder,0.111)
(VeriGen,0.100)
(VeriSeek,0.100)
(CodeV,0.000)
(Gemini,0.125)
(DeepSeek,1.000)
(GPT-5,0.500)
(Claude,0.333)
};

\addplot+[
    fill=teal!65!black,
    draw=black
]
coordinates {
(OriGen,0.111)
(RTLCoder,0.000)
(VeriGen,0.000)
(VeriSeek,0.000)
(CodeV,0.111)
(Gemini,0.875)
(DeepSeek,0.000)
(GPT-5,0.500)
(Claude,0.667)
};

\addplot+[
    fill=yellow!80!orange,
    draw=black
]
coordinates {
(OriGen,0.222)
(RTLCoder,0.000)
(VeriGen,0.000)
(VeriSeek,0.000)
(CodeV,0.222)
(Gemini,0.000)
(DeepSeek,0.000)
(GPT-5,0.000)
(Claude,0.000)
};

\legend{P/C,Not-Elab.,Partial,F-Mismatch}

\end{groupplot}
\end{tikzpicture}
\end{adjustbox}

\caption{FE-based robustness across models. The heatmaps report stage
rates and conditional yields on a shared 0--1 scale; the remaining
panels compare one-shot success, bounded success, synthesis eligibility,
and first-failure composition.}
\label{fig:metrics_panels}
\vspace{-4pt}
\end{figure*}
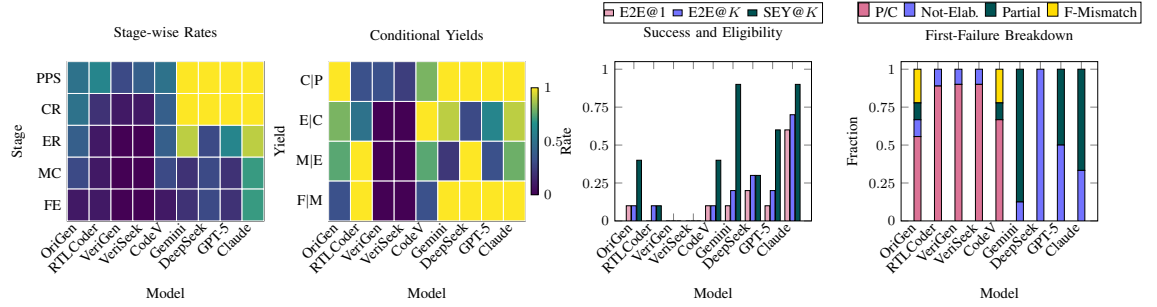

\subsection{Case Study: PID Controller}
\label{sec:pid}

The Genus table reports pre-layout logical-cell counts; next, we perform physical synthesis of the case study using Cadence Innovus to see the early evidence of ASIC place and route.  After place-and-route, the reference and Claude implementations contain 1,902 and 6,794 placed cells, respectively. Claude contains 5,190 combinational and 739 sequential cells, compared with 1,357 and 390 in the reference, reducing the sequential-cell ratio from 20.5\% to 10.9\% and exposing a wider combinational organization.

\begin{figure}[!t]
\centering
\begin{subfigure}[b]{0.47\columnwidth}
    \centering
    \includegraphics[width=\linewidth]{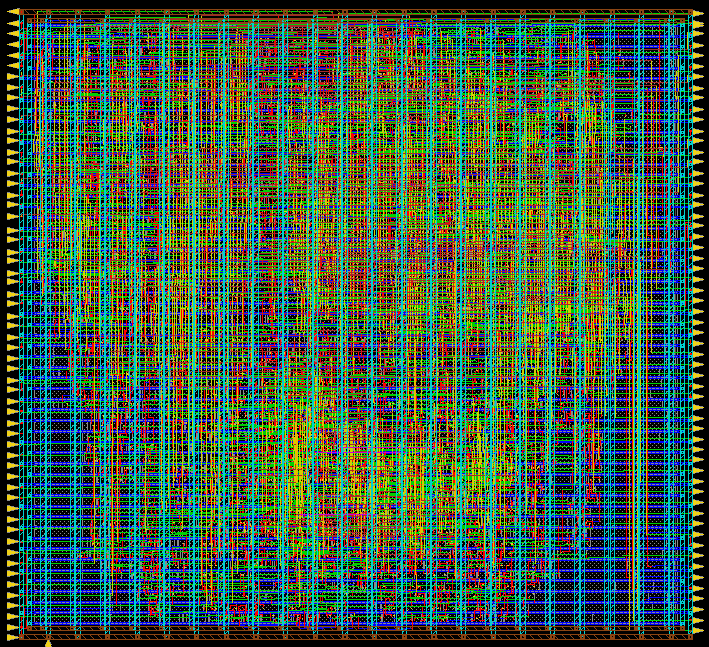}
    \caption{Baseline}
\end{subfigure}
\hfill
\begin{subfigure}[b]{0.47\columnwidth}
    \centering
    \includegraphics[width=\linewidth]{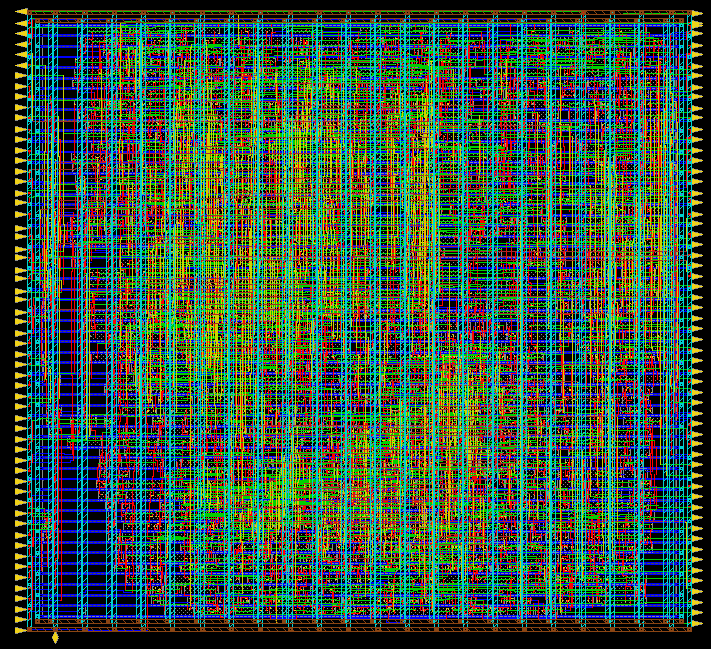}
    \caption{Claude}
\end{subfigure}
\vspace{1mm}
\caption{PID layouts after place-and-route in Cadence Innovus.}
\label{fig:pid_layout_compare}
\vspace{-2mm}
\end{figure}

Fig.~\ref{fig:pid_layout_compare} compares the two Innovus layouts.
Both designs complete the same place-and-route flow, but the
Claude-generated implementation occupies a larger area. The comparison illustrates why functional equivalence is not an implementation-quality metric: behaviorally equivalent RTL can expose substantially different logic organization and physical cost.

\subsection{Outcome of Evaluation Metrics}

Fig.~\ref{fig:metrics_panels} summarizes model behavior across the full benchmark. Panels~1-2 report stage-wise rates and conditional yields. The largest separations occur after Compilation, particularly at
Elaboration and Module Completeness, showing that syntactic validity
does not ensure coherent hierarchy or complete logic.

Panel~3 separates first-attempt robustness, bounded recovery, and
implementation survivability. E2E@1 reports complete front-end success
on the initial generation, while E2E@\(\!K\) additionally counts
designs recovered within the fixed refinement budget. The gap between
the two metrics therefore quantifies the benefit of bounded
specification clarification. SEY@\(\!K\) is broader and records
candidates that compile and elaborate sufficiently to enter synthesis,
even if they later fail MC or FE.

Panel~4 shows the earliest failure composition, separating
interface/compiler failures, non-elaboration, partial modules, and
functional mismatch.

\section{Discussion}
\label{sec:discussion}

Correctness remains essential, but the results show that it is
insufficient on its own. Many candidates fail before FE, while
candidates that pass can still map to substantially different hardware.
Evaluation should therefore distinguish structural validity,
behavioral correctness, and implementation efficiency rather than
collapsing them into one pass rate.

The implementation spread is most visible in datapath-intensive
designs, where coding style affects resource sharing, mux depth,
pipelining, and switching activity. Control-oriented blocks are more
stable, but they are not uniformly identical to their references. This
suggests that future RTL generators should receive hardware-aware
feedback in addition to compiler or equivalence diagnostics.

This study has four limitations. First, the suite uses public IP to
enable reference-equivalence checking, so overlap with model training
data cannot be excluded. Second, \(K=3\) measures bounded generation
robustness rather than unrestricted agentic repair. Third, the ASIC
study provides synthesis and one illustrative place-and-route case,
not full sign-off or silicon validation. Custom unseen specifications,
larger hierarchical subsystems, and cross-tool replication are natural
extensions. Fourth, each model-design pair is represented by one controlled
generation trajectory; the reported rates characterize this protocol
rather than the full stochastic output distribution of each model.

\section{Conclusion}

This paper presents an implementation-aware evaluation of
LLM-generated RTL for edge-hardware IP. Across nine models and ten
designs, compilation does not establish functional correctness, and
functional correctness does not establish implementation quality. The
proposed flow separates these concerns through staged validation,
synthesis-eligible yield, and FPGA/ASIC QoR evidence. For designers
adopting LLM-based RTL generation, the generated code should be judged not only by whether it works, but also by the hardware it becomes.

\bibliographystyle{IEEEtran}
\bibliography{dt-refs}

\begin{IEEEbiography}{Hepziba Susan}{\,}
is a graduate student in the Masters program in Integration Circuits and Systems at CentraleSupélec, Université Paris-Saclay, France. Her research interests include AI-assisted hardware design, RTL generation, VLSI, EDA, and hardware verification.
\end{IEEEbiography}

\begin{IEEEbiography}{Shivaranjani G.~R.}{\,}
is a final-year undergraduate student at Vellore Institute of Technology, Chennai, India. Her research interests include RTL design, functional verification, and AI-assisted electronic design automation.
\end{IEEEbiography}

\begin{IEEEbiography}{Malik Imran}{\,}
was a researcher at  Queen's University Belfast and is now an Assistant Professor at the University of Sharjah. His research interests include FPGA/ASIC design, hardware acceleration, and cryptographic
hardware.
\end{IEEEbiography}

\begin{IEEEbiography}{Muhammad Rashid}{\,}
is a Professor at Umm Al-Qura University, Makkah, Saudi Arabia. His research interests include embedded computing and AI-assisted electronic design automation. He has authored numerous publications in computer engineering.
\end{IEEEbiography}

\begin{IEEEbiography}{Sumathi Gokulanathan}{\,}
is an Assistant Professor with the School of Electronics Engineering, Vellore Institute of Technology (VIT), Vellore, India. Her research interests include VLSI design, electronic design automation,
and hardware security.
\end{IEEEbiography}

\begin{IEEEbiography}{Zain Ul Abideen}{\,}
is an Assistant Professor with the Department of ECE, University of Idaho, Moscow, ID, USA. His research interests include hardware security, security-aware electronic design automation (EDA), ASIC/FPGA design, post-quantum cryptography, and AI-assisted chip design. 
\end{IEEEbiography}

\end{document}